\documentclass[twocolumn,aps,prl,showpacs,amsmath,amssymb]{revtex4-1}
\usepackage{epsfig}
\usepackage{graphicx}
\usepackage{dcolumn}
\usepackage{bm}
\usepackage[colorlinks=true,dvipdfm]{hyperref}
\usepackage{multirow}

\begin{document}
\title{Kibble-Zurek scaling in the Yang-Lee edge singularity}
\author{Shuai Yin$^{1,2}$}
\author{Guang-Yao Huang$^{3}$}
\author{Chung-Yu Lo$^{1}$}
\author{Pochung Chen$^{1}$}
\affiliation{$^{1}$Department of physics, National Tsing Hua University, Hsinchu 30013, Taiwan}
\affiliation{$^{2}$Institute for Advanced Study, Tsinghua University, Beijing, 100084, P. R. China}
\affiliation{$^{3}$Department of Electronics and Key Laboratory for the Physics and Chemistry of Nanodevices, Peking University, Beijing 100871, P. R. China}
\date{\today}

\begin{abstract}
We study the driven dynamics across the critical points of the Yang-Lee edge singularities (YLESes) in a finite-size quantum Ising chain with an imaginary symmetry-breaking field. In contrast to the conventional classical or quantum phase transitions, these phase transitions are induced by tuning the strength of the dissipation in a non-Hermitian system and can occur even at finite size. For conventional phase transitions, universal behaviors in driven dynamics across critical points are usually described by the Kibble-Zurek mechanism, which states that the scaling in dynamics is dictated by the critical exponents associated with one critical point and topological defects will emerge after the quench. While the mechanism leading to topological defects breaks down in the YLES, we find that for small lattice size, the driven dynamics can still be described by the Kibble-Zurek scaling with the exponents determined by the $(0+1)$-dimensional YLES. For medium finite size, however, the driven dynamics can be described by the Kibble-Zurek scaling with two sets of critical exponents determined by both the $(0+1)$-dimensional and the $(1+1)$-dimensional YLESes.

\end{abstract}
\pacs{03.67.Mn, 64.60.De, 64.60.Ht, 64.70.Tg}
\maketitle

The Kibble-Zurek mechanism~\cite{Kibble,Zurek} describes universal scaling behavior in the driven critical dynamics in a variety of systems, ranging from classical to quantum phase transitions~\cite{Dz,Pol}. It separates the whole driven process into three stages: two adiabatic stages and one impulse stage. In the adiabatic region, the relaxation rate is larger than the transition rate and the system evolves along the instantaneous equilibrium state; while in the impulse stage, the relaxation rate is smaller than the transition rate because of the critical slowing down, and thus the system falls out of equilibrium essentially. Furthermore, the Kibble-Zurek scaling (KZS)~\cite{Kibble,Zurek} shows that only the equilibrium critical exponents are needed to characterise the dynamic scaling behavior. All these exponents belong to one set, which is determined by the renormalization group flow near the critical point~\cite{Dz,Pol,Zhongfts}. According to the KZS, the external driving will induce an effective correlation length, which divides the system into different domains. The domain walls will form topological defects, whose number can be scaled by the driving rate~\cite{Dz,Pol}. Additionally, for a finite-size system, the system size also becomes an scaling variable~\cite{cdg,Yin,Huang}. The KZS has been verified numerically and experimentally in both classical and quantum phase transitions~\cite{qkz1,qkz2,qkz3,NDAntunes,qkz4,qkz5,BDamski,qkz7,qkz8,qkz9,Chandran,Ulm,Pyka,Navon,Clark,Hu,CWLiu1,CWLiu2}.

On the other hand, Yang and Lee~\cite{Yang,Lee} paved the way to understand phase transitions by analysing the zeros of the partition function in the complex plane of a symmetry-breaking field. It was shown that singular behaviors exist not only at the critical point with a vanishing symmetry-breaking field but also near the edge of the Lee-Yang zeros, where the applied symmetry-breaking field is purely imaginary~\cite{Kortman}. The latter case is often referred to as the Yang-Lee edge singularity (YLES) and can be cast to a critical theory characterized by the Landau-Ginzburg action of a scalar field with an imaginary cubic coupling~\cite{Fisher}. Although the YLES occurs in the complex parameter space, its critical properties can be detected in experiments~\cite{Binek,WeiLiu,PengLiu}.

While there are many exotic scaling behaviors in YLES such as the divergence of the order parameter and negative correlation-length exponent in low dimensions~\cite{Fisher}, to the best of our knowledge, however, the nonequilibrium properties of the YLES has rarely been investigated.
Furthermore, the quantum YLES provides a prototype to study a class of dissipative phase transitions, which is characterised by the spontaneous parity-time (PT) symmetry breaking~\cite{Moiseyevbook}. Different from usual quantum phase transitions which occur by tuning a parameter in the Hermitian Hamiltonian, dissipative phase transitions are induced by changing the strength of the dissipation~\cite{Weiss}. Recently driven-dissipative open quantum systems have attracted much attention as they offer a promising route of quantum computations or state engineering~\cite{Weiss,Diehl,Diehl2,Moiseyevbook,Bender}. In addition, PT symmetry-breaking phase transitions have been observed in both optic and ultracold atoms systems~\cite{ExpPeng,ExpZhen,ExpFeng,ExpZhang,LuoLe}. These call for a study on the nonequilibrum behavior near their phase transitions. Some questions then arise: How to describe the driven dynamics across such phase transitions which exhibit YLES? Is the Kibble-Zurek mechanism still applicable? If the answer is yes, is there any new ingredient in such KZS?

To answer these questions, we study the driven dynamics across the critical point of YLESes in a finite-size quantum Ising chain with an imaginary symmetry-breaking field~\cite{Uzelac}. We confirm that the KZS is applicable but there are some features which are quite different from the KZS in ordinary phase transitions. In particular, we show that while the mechanism leading to topological defects breaks down, for small size system the driven dynamics is still described by the KZS with $(0+1)$D critical exponents. For the medium size system, however, the driven dynamics can be described by the KZS with both $(0+1)$D and $(1+1)$D critical exponents. The experimental feasibility of the KZS in the YLES is then discussed.


\emph{Static properties of the YLES}--- We begin our study with the quantum Ising chain in an imaginary longitudinal field~\cite{Uzelac}. The Hamiltonian reads
\begin{equation}
\mathcal{H}=-\sum_{n=1}^{L}\sigma_n^z\sigma_{n+1}^z-\lambda\sum_{n=1}^{L}\sigma_n ^x-{\rm i}h\sum_{n=1}^{L}\sigma_n^z,
\label{HIsing}
\end{equation}
where $\sigma_n^z$ and $\sigma_n^x$ are the Pauli matrices in $z$ and $x$ direction, respectively, at site $n$, $\lambda$ is the transverse-field, $h$ is longitudinal-field, and $L$ is the lattice size. The critical point of the ordinary ferromagnetic-paramagnetic phase transition is $\lambda_{c}=1$ and $h=0$~\cite{Sachdev} while there are critical points for the YLES at $(\lambda_{\rm YL}^L, h_{\rm YL}^L)$ when $\lambda>\lambda_{c}$~\cite{Fisher} (The superscript capital $L$ indicates the lattice size instead of an exponent).
Although $\mathcal{H}$ is non-Hermitian, the appearance of the YLES corresponds to the vanishment of the energy gap~\cite{Gehlen,supmate}, similar to the ordinary quantum critical phenomena occurring in the Hermitian system~\cite{Sachdev}. One can also define an order parameter: $M\equiv|{\rm Re}[\langle\Psi^*|\sigma^z|\Psi\rangle/\langle\Psi^*|\Psi\rangle]|$~\cite{Uzelac,Gehlen}. For a fixed $\lambda (\lambda > \lambda_c)$, when $h<h_{\rm YL}^L$, the real part of model~(\ref{HIsing}) dominates and the energy spectra are real~\cite{supmate}. Since these spectra are adiabatically connected with those for $h=0$, the system is in a paramagnetic phase with $M=0$. When $h>h_{\rm YL}^L$, the dissipative part in model~(\ref{HIsing}) plays significant roles. As a result, energy spectra become conjugate pairs~\cite{Gehlen,supmate}. It has been shown that the latter phase is a ferromagnetic phase with $M\neq0$~\cite{Kortman}. Moreover, it has been demonstrated that the equilibrium singular behaviors near the critical point of the YLES can be described by the usual critical exponents~\cite{Fisher}. For instance, $\beta_0=\beta_1=1$, $\nu_0=-1$, $\delta_0=-2$, $\nu_1=-5/2$, $\delta_1=-6$ and the dynamic exponents $z_0=z_1=1$~\cite{Fisher,Uzelac,Gehlen}. (The subscript indicates the space dimension). We note that $M$ diverges at the YLES because $\delta$ is negative.

\begin{figure}
  \centering
   \includegraphics[bb= 0 0 162 105, clip, scale=1.1]{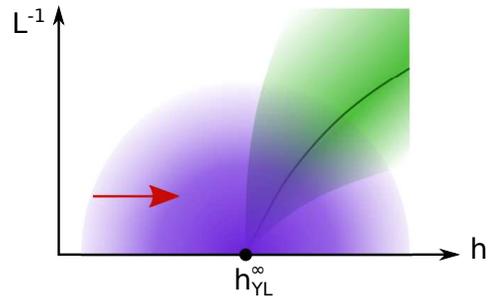}
   \caption{\label{drivenregion}(Color online) Critical regions near critical points of YLESes. Critical points of finite-size YLESes link up into a critical curve (Solid green curve), which ends at the critical point of the infinite-size YLES $h_{\rm YL}^\infty$. The red arrow indicates the direction of changing $h$ in KZS. For a small size system, the evolution will cross the critical region (Green) described by the $(0+1)$D theory of the finite-size YLES; while for the system with a medium lattice-size, the evolution will cross the critical region (Blue) described by the $(1+1)$D theory of the infinite-size YLES and the overlap between these two critical regions. We show that the driven dynamics in the overlap region can be described by the KZS according to both the $(0+1)$D and $(1+1)$D critical theories of YLESes.}
\end{figure}

Different from usual phase transitions which only occur in the thermodynamic limit, the YLES can appear even at finite sizes~\cite{Kortman,Gehlen}. The YLES near $h_{\rm YL}^L$ in model~(\ref{HIsing}) belongs to the $(0+1)$D universality class, while the YLES near $h_{\rm YL}^\infty$ belongs to the $(1+1)$D universality class~\cite{Fisher}. Furthermore, it has been shown that~\cite{Gehlen}
\begin{equation}
h_{\rm YL}^L-h_{\rm YL}^\infty=C(\lambda)L^{-\frac{\beta_1\delta_1}{\nu_1}},
\label{scalingh}
\end{equation}
in which $C(\lambda)$ is a dimensionless function. Two deductions thus can be obtained as sketched in Fig.~\ref{drivenregion}: (i) there must be an overlap critical region in which both the $(0+1)$D YLES and the $(1+1)$D YLES play indispensable roles; (ii) the critical region for the $(0+1)$D YLES must shrink as $L$ increases, and when $L\rightarrow\infty$, this region becomes a point.

\emph{KZS for small-size systems}--- We first study the KZS for model (\ref{HIsing}) with a small size system, whose critical region is the green region in Fig.~\ref{drivenregion}. The critical properties in this region are described by the $(0+1)$D critical theory. We consider the case for changing $h$ as $h=h_0+R_ht$, while $\lambda (> \lambda_c)$ is fixed. Since $h_0$ is chosen to be far away from the YLES, it is irrelevant~\cite{YinChen}. Similar to the KZS in ordinary phase transitions~\cite{Dz,Pol}, when $|h-h_{\rm YL}^L|>R_h^{{\beta_0 \delta_0}/{\nu_0 r_0}}$ ($r_0=z_0+\beta_0\delta_0/\nu_0$), the relaxation rate, $|h-h_{\rm YL}^L|^{\nu_0 z_0/\beta_0 \delta_0}$, is larger than the transition rate, $R_h/|h-h_{\rm YL}^L|$, and the evolution is in the adiabatic stage; and when $|h-h_{\rm YL}^L|<R_h^{{\beta_0 \delta_0}/{\nu_0 r_0}}$, the relaxation rate is smaller than the transition rate and the evolution is in the impulse stage~\cite{qkz1,qkz2,qkz3,NDAntunes}. Therefore, the dynamics of $M$ near $h_{\rm YL}^L$ should still satisfy the KZS~\cite{Chandran,Zhongfts}
\begin{equation}
M(h-h_{\rm YL}^L,R_h)=R_h^{\frac{\beta_0}{\nu_0 r_0}}f_a[(h-h_{\rm YL}^L)R_h^{-\frac{\beta_0 \delta_0}{\nu_0 r_0}}],
\label{scaling2}
\end{equation}
in which $f_a$ is an analytical scaling function (similar definitions will always be implied). Equation (\ref{scaling2}) is applicable when the impulse region is embedded in the critical region of the $(0+1)$D critical point~\cite{Silvi}. Otherwise, the information, which is not controlled by the $(0+1)$D critical theory, can be brought into the driven dynamics. Moreover, although we focus on the small-size system, there is no finite-size correction in Eq.~(\ref{scaling2}), since $L$ is irrelevant in this $(0+1)$D YLES. Additionally, for the driven dynamics in ordinary phase transitions, topological defects emerge after impulse stage since the whole lattice of the system is divided by the driven-induced length scale; in the $(0+1)$D YLES, however, the topological defects are not well-defined, since the lattice size can be microscopically small.


\emph{KZS in the overlap region}--- For medium sizes, the critical regions for the $(0+1)$D and $(1+1)$D overlap with each other, as shown in Fig.~\ref{drivenregion}. In this overlap region, besides Eq.~(\ref{scaling2}), the dynamic scaling should satisfy the $(1+1)$D KZS with finite-size corrections being considered. Similar to the usual finite-size KZS~\cite{Yin,cdg}, the scaling form of $M$ reads
\begin{equation}
M(h-h_{\rm YL}^\infty,R_h,L)=R_h^{\frac{\beta_1}{\nu_1 r_1}}f_b[(h-h_{\rm YL}^\infty)R_h^{-\frac{\beta_1\delta_1}{\nu_1 r_1}},L^{-1}R_h^{-\frac{1}{r_1}}],
\label{scaling1}
\end{equation}
in which $r_1=z_1+\beta_1\delta_1/\nu_1$. Equation~(\ref{scaling1}) is applicable in the $(1+1)$D critical region (Blue region in Fig.~\ref{drivenregion}). Since both $f_a$ and $f_b$ are analytical functions for any finite $L$ and $R_h$, both Eq.~(\ref{scaling2}) and Eq.~(\ref{scaling1}) should be applicable in the overlap region. So, there must be some latent scaling properties for $f_a$ and $f_b$.

To explore these properties, we start from $f_b$. When $R_h\rightarrow0$ and $h\rightarrow h_{\rm YL}^L$, $M$ diverges as $M\sim (h-h_{\rm YL}^L)^{1/\delta_0}$ according to the $(0+1)$D static scaling theory~\cite{supmate}. However, $h_{\rm YL}^L$ is not an explicit variable in $f_b$. To expose this divergence, we substitute Eq.~(\ref{scalingh}) into Eq.~(\ref{scaling1}) and obtain
\begin{equation}
M(h-h_{\rm YL}^L,R_h,L)=R_h^{\frac{\beta_1}{\nu_1 r_1}}f_c[(h-h_{\rm YL}^L)R_h^{-\frac{\beta_1\delta_1}{\nu_1 r_1}},L^{-1}R_h^{-\frac{1}{r_1}}].
\label{scaling2a}
\end{equation}
Comparing Eq.~(\ref{scaling2a}) with Eq.~(\ref{scaling2}), one finds that the scaling function $f_c(A,B)$ satisfy
\begin{equation}
f_c(A,B)=(B^{-r_1})^{\frac{\beta_0}{\nu_0 r_0}-\frac{\beta_1}{\nu_1 r_1}}f_d[A(B^{-r_1})^{\frac{\beta_1 \delta_1}{\nu_1 r_1}-\frac{\beta_0\delta_0}{\nu_0 r_0}}].
\label{scaling3}
\end{equation}
Thus, Eq~(\ref{scaling3}) provides a constraint, which make the explicit scaling variable $L$ in $f_b$ ($f_c$) behave like a dimensionless parameter and Eq.~(\ref{scaling2}) is restored.

As $L\rightarrow\infty$, the driven dynamics must be described by the $(1+1)$D KZS theory, i.e., Eq.~(\ref{scaling1}) with $L\rightarrow\infty$. However, at first glance, by taking $L\rightarrow\infty$ in Eq.~(\ref{scaling2}), one obtains $M(h,R_h)=R_h^{\beta_0/\nu_0 r_0}f_a[(h-h_{\rm YL}^\infty)R_h^{-\beta_0 \delta_0/\nu_0 r_0}]$. This is apparently incorrect. The reason is that the critical region of the $(0+1)$D critical point shrinks to a point as $L\rightarrow\infty$. So, for any finite driving rate, the impulse region of the $(0+1)$D KZS is broader than the $(0+1)$D critical region. As a consequence, Eq.~(\ref{scaling2}) is not applicable anymore when $L\rightarrow\infty$.



\begin{figure}
  \centerline{\epsfig{file=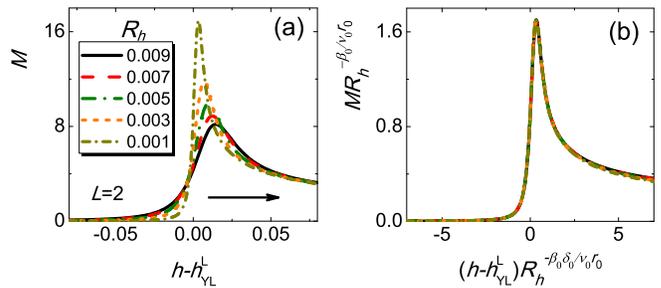,width=1.0\columnwidth}}
  \caption{\label{smallL} (Color online) Under increasing $h$ with fixed $\lambda=5$, the curves of $M$ versus $h-h_{\rm YL}^{L}$ ($h_{\rm YL}^{L}=2.933353$) for fixed $L=2$ in (a) match with each other in (b) when $M$ and $h-h_{\rm YL}^{L}$ are rescaled by the $(0+1)$D exponents. The arrow points the direction of changing $h$. $h_0$ is chosen as $h_0\equiv h_{\rm YL}^{L}-1$.}
\end{figure}

\emph{Numerical results}--- We numerically solve Schr\"{o}dinger's equation for model (\ref{HIsing})~\cite{supmate}. For small sizes, Fig.~\ref{smallL} shows the evolution of $M$ for $h=R_ht+h_0$ with a fixed $\lambda$ ($\lambda>\lambda_c$). First, we find that the divergence of $M$ at $h_{\rm YL}^L$ is rounded by the external driving. Second, in Figs.~\ref{smallL}(b), after rescaling $M$ and $h$ with $R_h$ by using the $(0+1)$D exponents, we find that the rescaled curves match with each other in the vicinity of $h_{\rm YL}^L$, confirming Eq.~(\ref{scaling2}).

\begin{figure}
  \centerline{\epsfig{file=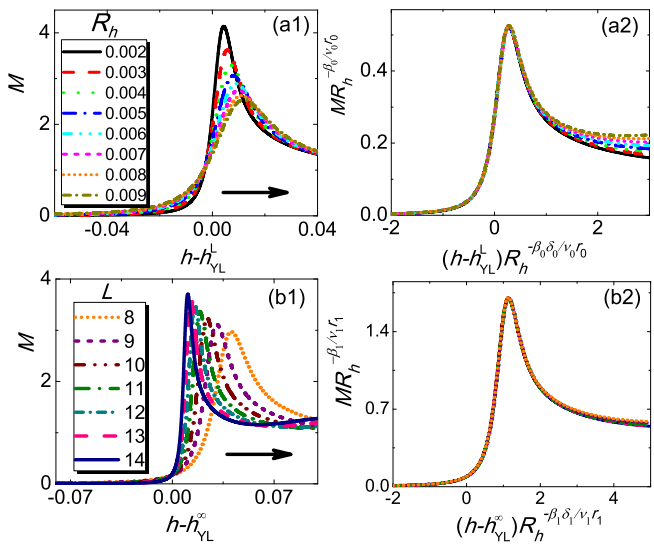,width=1.0\columnwidth}}
  \caption{\label{dbscaling} (Color online) Under increasing $h$ with fixed $\lambda=5$, the curves of $M$ versus $h-h_{\rm YL}^{L}$ ($h_{\rm YL}^{L}=2.309176$) for fixed $L=10$ in (a1) match with each other in (a2) when $M$ and $h-h_{\rm YL}^{L}$ are rescaled by the $(0+1)$D exponents; for comparison, the curves of $M$ versus $h-h_{\rm YL}^\infty$ for fixed $LR_h^{1/r_1}=2.001689$ in (b1) match with each other in (b2) when $M$ and $h-h_{\rm YL}^\infty$ are rescaled by the $(1+1)$D exponents. The arrows point the directions of changing $h$. $h_0$ is chosen as $h_0\equiv h_{\rm YL}^{\infty}-1$ for both (a) and (b).}
\end{figure}

For medium sizes, we compare the driven dynamics for fixed $L$ and fixed $LR_h^{1/r_1}$. In Figs.~\ref{dbscaling}(a1) and~\ref{dbscaling}(a2), $L$ is fixed and the $(0+1)$D critical exponents are employed to calculate the rescaled variables. Similar to Fig.~\ref{smallL}, after rescaling $M$ and $h$ with $R_h$, we find that the rescaled curves match with each other in the vicinity of $h_{\rm YL}^L$, confirming Eq.~(\ref{scaling2}). In contrast, in Figs.~\ref{dbscaling}(b1) and~\ref{dbscaling}(b2), $LR_h^{1/r_1}$ is fixed and the rescaled values are calculated according to the $(1+1)$D theory. We find that the rescaled curves collapse onto each other according to Eq.~(\ref{scaling1}). Thus, we conclude that the driven critical dynamics near the critical point of YLES can be described by the KZS in both $(0+1)$D and $(1+1)$D. Comparing Fig.~\ref{dbscaling}(a2) with Fig.~\ref{dbscaling}(b2), we find that after the peaks, the collapse in Fig.~\ref{dbscaling}(b2) is much better than that in Fig.~\ref{dbscaling}(a2). The reason is that the critical region for the $(1+1)$D YLES is broader than the critical region for the $(0+1)$D YLES, as shown in Fig.~(\ref{drivenregion}). Moreover, comparing Figs.~\ref{smallL} and \ref{dbscaling} (a), we find that the collapse region for the rescaled curves becomes smaller as the lattice size increases. This indicates that the regime, in which Eq.~(\ref{scaling2}) is applicable, shrinks for larger $L$.

\begin{figure}
  \centerline{\epsfig{file=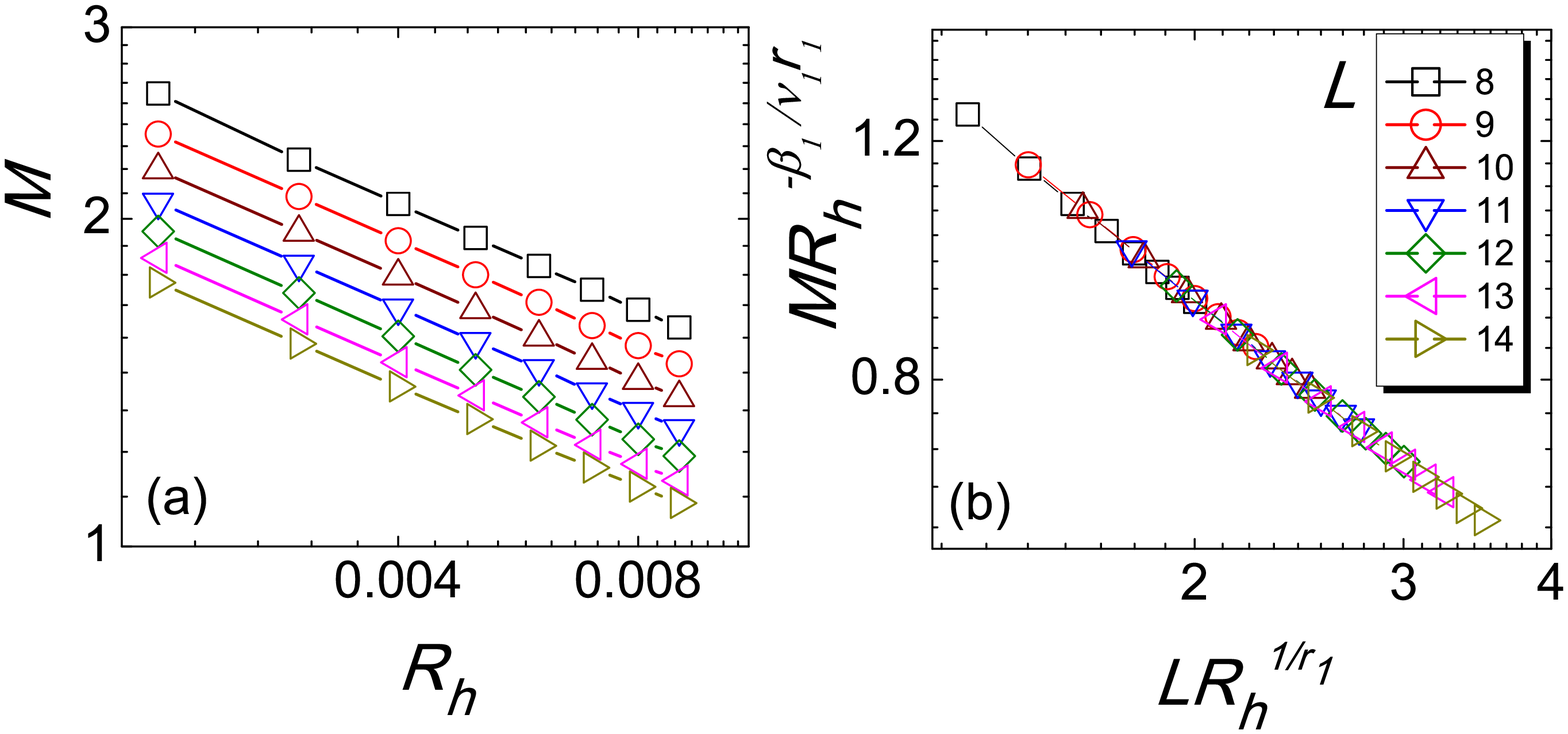,width=1.0\columnwidth}}
  \caption{\label{scarel} (Color online) Under changing $h$ with fixed $\lambda=5$, (a) $M$ at $h_{\rm YL}^L$ versus $R_h$ for different lattice sizes; and (b) the collapse of curves of the rescaled $M$ at $\lambda_{\rm YL}^L$ versus the rescaled $L$. Double-logrithmic scales are used in both (a) and (b).}
\end{figure}

To investigate the relation between Eqs.~(\ref{scaling2}) and (\ref{scaling1}), we extract the order parameters at $h_{\rm YL}^L$ for various $R_h$. First, Fig.~\ref{scarel}(a), plotted on the double-logarithmic scale, shows that for different lattice sizes, the curves of $M$ versus $R_h$ are almost parallel straight lines, whose slopes are between $-0.334$ and $-0.324$, agreeing with the theoretical value of $\beta_0/\nu_0 r_0=-1/3$. Second, we plot in Fig.~\ref{scarel}(b) the rescaled order parameter as the function of the rescaled lattice size with the $(1+1)$D critical exponents as input. The rescaled curves collapse onto one single curve. Thus, Eq.~(\ref{scaling2a}) is confirmed and this rescaled curve is just the scaling function $f_c(0,B)$. Third, as shown in Fig.~\ref{scarel}(b), by plotting the rescaled curve in double-logarithmic scale, one finds that $f_c(0,B)$ itself is a power function, whose exponent is fitted to be $-0.731$. This exponent is close to the theoretical value of $r_1(\beta_0\delta_0/\nu_0 r_0-\beta_1\delta_1/\nu_1 r_1)\simeq-0.7333$, confirming Eq.~(\ref{scaling3}).

\emph{Discussion}--- Here we give an intuitive understanding of the KZS in YLES. The applicability of the KZS in conventional quantum phase transitions depends on two facts: one is that besides the ground state, low-energy excited states also exhibit universal scaling properties; the other is that the probability of occupying the excited states scales with the driving rate. We expect that the KZS in the YLES will satisfy both conditions. Consider now the case for slow driving such that only two energy levels get involved (See Figs. 1 in supplemental material~\cite{supmate}). In the static case, $\partial E/\partial h$ diverges for each single level at the YLES. However, under an external driving, the mixed contributions from these two involved levels may cancel the divergence since their derivatives have opposite sign. This explains the rounding of $M$ for changing $h$ with a finite rate. For different driving rate, the probability of occupying different levels can be scaled by the driving rate, resulting in the KZS for $M$. The difference is that the topological defects, which are always used to characterize the driven dynamic in conventional phase transitions, are not well-defined in the $(0+1)$D YLES.

\emph{Experimental feasibility}--- Here, we discuss possible experimental approaches to detect the KZS in the YLES. The first approach is based on the fact that the YLES has been found in FeCl$_2$~\cite{Binek}. The corresponding KZS can be examined therein. The second one is based on the method, proposed in~\cite{WeiLiu,PengLiu}, of detecting Lee-Yang zeros by measuring quantum coherence of a probe spin in an Ising bath. The YLES for the density of Lee-Yang zeros has been observed in~\cite{WeiLiu} and the location of the YLES has been estimated experimentally~\cite{PengLiu}. The KZS of the YLES can be realized by imposing a time-dependent term in the Hamiltonian therein. Since the density of Lee-Yang zeros is proportional to $M$~\cite{Kortman}, the KZS for the density of Lee-Yang zeros should satisfy the same scaling theory proposed here. For experimental convenience, we have also considered the cases for changing $\lambda$ and real longitudinal field $j$ near the YLES (see supplemental material~\cite{supmate}). Besides, it is also expected that our scaling theory can be examined in the PT symmetry-breaking transitions~\cite{ExpPeng,ExpZhen,ExpFeng,ExpZhang,LuoLe}. For instance, we show in \cite{supmate} that the KZS for a PT symmetry-breaking phase transition, observed in a recent experiment~\cite{LuoLe}, satisfies Eq.~(\ref{scaling2}).

\emph{Summary}--- In summary, we have studied the driven dynamics in the YLES. For the $(0+1)$D YLES, we have shown that although the topological defects is not well-defined, the KZS can still be applied to describe the driven dynamics. Additionally, in the overlap critical regions between the $(0+1)$D and $(1+1)$D YLES, we have found that the driven critical dynamics can be described by the KZS according to both the $(0+1)$D and $(1+1)$D critical theories, although their critical exponents are different. We have also explored the relation between dynamic scaling functions of the KZS in the $(0+1)$D and $(1+1)$D theory. Possible experimental realizations has then been discussed.

Our investigations on the KZS in YLES can be readily generalized into other systems, including the classical YLES~\cite{Fisher,Janssen}, and other PT symmetry-breaking transitions. In addition, our KZS for the overlap critical region is also applicable in other systems. For example, the two-dimensional quantum Ising model at finite temperatures exhibits both classical and quantum phase transition~\cite{Sachdev}, and its phase diagram is similar to Fig.~\ref{drivenregion}. The KZS therein should obey similar scaling theories. Besides, since the PT symmetry phase transition can occur in optic systems~\cite{ExpPeng,ExpZhen,ExpFeng,ExpZhang}, it is appealing to explore whether the KZS is also applicable therein.

We wish to thank F. Zhong, Z. Wang, Z. Yan, S. Zhang, and R. He for their helpful discussions. We acknowledge the support by Ministry of Science and Technology (MOST) of Taiwan through Grant No. 104-2628-M-007-005-MY3. We also acknowledge the support from the National Center for Theoretical Science (NCTS) of Taiwan.

\begin{widetext}
\clearpage

\section{ Supplemental material }

\subsection{\uppercase\expandafter{\romannumeral1}. The Yang-Lee edge singularity at finite size}
\emph{Energy spectra}--- To illustrate the Yang-Lee edge singularity (YLES) for finite-size systems~\cite{Gehlensup}, we show the lowest two eigenvalues for model (1) in the main text. In Fig.~\ref{kk}, one finds that for fixed $\lambda$, when $h<h_{\rm YL}^L$, the spectra are real; when $h>h_{\rm YL}^L$, the spectra form conjugate pairs; and exactly at $h_{\rm YL}^L$, the gap between the lowest two eigenvalues vanishes. For comparison, in Fig.~\ref{kk1}, one finds that for fixed $h$, when $\lambda>\lambda_{\rm YL}^L$, the spectra are real; when $\lambda<\lambda_{\rm YL}^L$, the spectra form conjugate pairs; and at $\lambda_{\rm YL}^L$, the gap vanishes.

\begin{figure}[htb]
  \centering
   \centerline{\epsfig{file=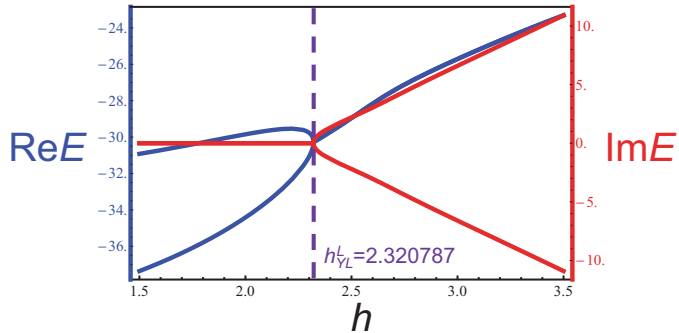,width=0.5\columnwidth}}
   \caption{\label{kk}(Color online) The lowest two levels for $\lambda=5$ and $L=8$. The gap vanishes at $h_{\rm YL}^{L}=2.320787$.}
\end{figure}

\begin{figure}[htb]
  \centering
   \centerline{\epsfig{file=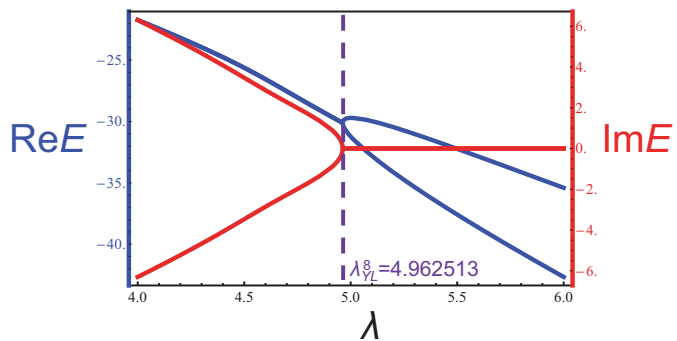,width=0.5\columnwidth}}
   \caption{\label{kk1}(Color online) The lowest two levels for $h=2.292475$ and $L=8$. The gap vanishes at $\lambda_{\rm YL}^{L}=4.962513$.}
\end{figure}

\emph{Estimation of the critical point of the YLES}--- As shown in Fig.~\ref{findcritical}, we estimate the critical point of the YLES by determining the position at which the order parameter diverges~\cite{Gehlensup,Uzelacsup,GarWei}. Different from usual phase transitions, in which $M$ vanishes at the critical point, here, $M$ tends to infinity when $h$ ($\lambda$) tends to $h_{\rm YL}^L$ ($\lambda_{\rm YL}^L$) from the side of the ferromagnetic phase; while $M=0$ when $h$ ($\lambda$) tends to $h_{\rm YL}^L$ ($\lambda_{\rm YL}^L$) from the side of the paramagnetic phase. In Table~\ref{tab1}, we list values of $h_{\rm YL}^L$ and $\lambda_{\rm YL}^L$ with fixed $\lambda$ and $h$, respectively, for different $L$.
\begin{figure}[htb]
  \centerline{\epsfig{file=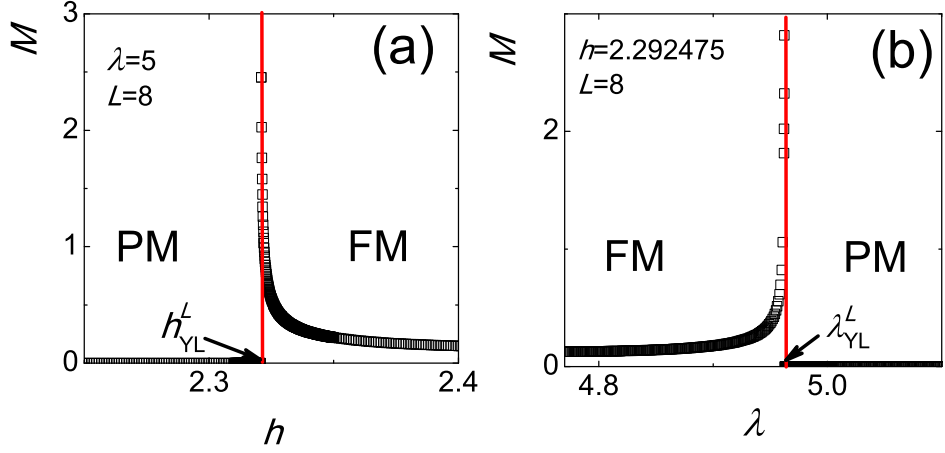,width=0.7\columnwidth}}
  \caption{\label{findcritical} (Color online) (a) Estimations of $h_{\rm YL}^L$ for fixed $\lambda$; (b) Estimation of $\lambda_{\rm YL}^L$ for fixed $h$. }
\end{figure}
\begin{table}[htbp]
  \centering
  \caption{The critical points of the YLES for various lattice sizes}
    \begin{tabular}{c l| c | c }
    \hline
    \hline
    \multicolumn{2}{c|}{$L$} & $h_{\rm YL}^L$ for $\lambda=5$ & $\lambda_{\rm YL}^L$ for $h=2.292475$  \\
    \hline
          & $8$     & $2.320787$   & $4.962513$    \\
          & $9$     & $2.313911$  & $4.971616$  \\
          & $10$    & $2.309176$  & $4.977885$   \\
          & $11$    & $2.305794$  & $4.982363$   \\
          & $12$    & $2.303305$  & $4.985659$  \\
          & $13$    & $2.301426$  & $4.988147$   \\
          & $14$    & $2.299978$  & $4.990065$  \\
    \hline
    \end{tabular}%
  \label{tab1}%
\end{table}%

\emph{Finite-size scaling of the critical point of the YLES}--- Here we show the finite-size scaling of the critical point of the finite-size YLES~\cite{Gehlensup}. For fixed $\lambda$ ($=5$), we plot in Fig~\ref{FSSYLE}(a) the difference between $h_{\rm YL}^L$ and $h_{\rm YL}^\infty(=2.292475)$ as a function of $L$. Power fitting shows that this curve satisfies $(h_{\rm YL}^L-h_{\rm YL}^\infty)\propto L^{-2.366}$, approximately agree with Eq.~(2) in which $\beta_1\delta_1/\nu_1=12/5$. For comparison, with fixed $h$ ($=2.292475$), we plot in Fig~\ref{FSSYLE}(b) the difference between $\lambda_{\rm YL}^L$ and $\lambda_{\rm YL}^\infty(=5)$ as a function of $L$. Power fitting shows that this curve satisfies $(\lambda_{\rm YL}^L-\lambda_{\rm YL}^\infty)\propto L^{-2.370}$. This indicates that $\lambda$ has the same critical dimension with $h$~\cite{Gehlensup}. Here we only consider the leading term. The higher order corrections have been discussed in Ref.~\onlinecite{Gehlensup}.
\begin{figure}[htb]
  \centerline{\epsfig{file=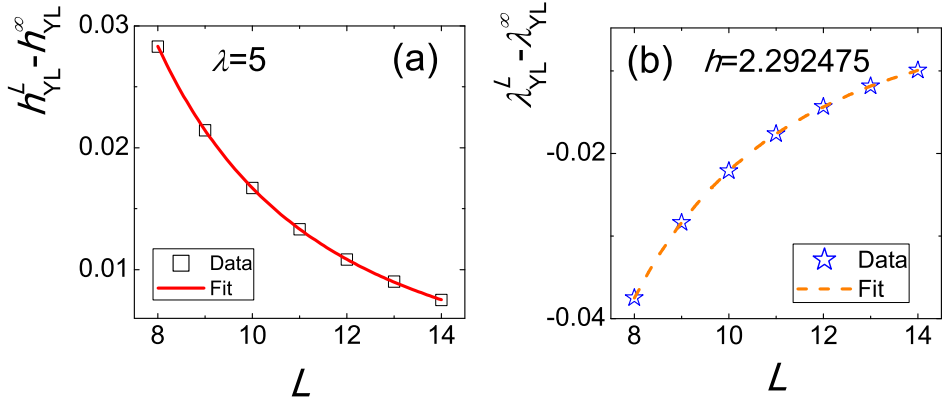,width=0.7\columnwidth}}
  \caption{\label{FSSYLE} (Color online) (a) Fitting of $(h_{\rm YL}^L-h_{\rm YL}^\infty)$ versus $L$ for fixed $\lambda$. (b) Fitting of $(\lambda_{\rm YL}^L-\lambda_{\rm YL}^\infty)$ versus $L$ for fixed $h$.}
\end{figure}

\subsection{\uppercase\expandafter{\romannumeral2}. Numerical method}
To verify the scaling theory, we solve directly the Schr\"{o}dinger equation of the Hamiltonian~(1) in the main text by using the finite difference method in the time direction with periodic boundary condition. The time interval is chosen as $5\times10^{-5}$. Smaller intervals have been checked to produce no appreciable changes. The initial wavefunction is chosen as the ground state wavefunction, which is obtained by the exact diagonalization, for the initial parameter. After each step of the evolution, the wavefunction is normalized as $\langle\Psi|\Psi\rangle=1$. This normalization process will not change the expectation value of $M$.

\subsection{\uppercase\expandafter{\romannumeral3}. The KZS for changing the transverse-field $\lambda$}
Besides changing $h$, one can also change $\lambda$ to cross the critical point of the YLES. It has been proved that there is only one relevant direction in the parameter space of the YLES~\cite{Fishersup,Gehlensup}. Therefore, the KZS for changing $\lambda$ is exactly the same as that for changing $h$. This feature is different from the KZS in ordinary classical and quantum phase transitions, in which the relevant exponents of the KZS are usually different for changing different parameters.

We consider the driving $\lambda=\lambda_0+R_\lambda t$ with $\lambda_0$ being far from the YLES. In Figs.~\ref{dbscalingl}(a1) and~\ref{dbscalingl}(a2), $L$ is fixed and the $(0+1)$D critical exponents are employed to calculate the rescaled variables. After rescaling $M$ and $(\lambda-\lambda_{\rm YL}^L)$ with $R_\lambda$, we find that the rescaled curves match with each other in the vicinity of $\lambda_{\rm YL}^L$, confirming Eq.~(3) with $h$ and $R_h$ being replaced by $\lambda$ and $R_\lambda$, respectively. In Figs.~\ref{dbscalingl}(b1) and~\ref{dbscalingl}(b2), $LR_\lambda^{1/r_1}$ is fixed and the rescaled values are calculated according to the $(1+1)$D theory. We find that the rescaled curves collapse onto each other according to Eq.~(4) with $h$ and $R_h$ being replaced by $\lambda$ and $R_\lambda$, respectively. Thus, similar to the case of changing $h$, we conclude that the critical dynamics under changing $\lambda$ near the critical point of the YLES of model~(1) can be described by the KZS for both $(0+1)$D and $(1+1)$D. Comparing Fig.~\ref{dbscalingl}(a2) with Fig.~\ref{dbscalingl}(b2), we find that on the left-hand side of the peaks, the collapse in (b2) is much better than that in (a2). The reason is that the critical region for the infinite-size YLES is broader than the critical region for the finite-size YLES, similar to the case of changing $h$.
\begin{figure}[htb]
  \centerline{\epsfig{file=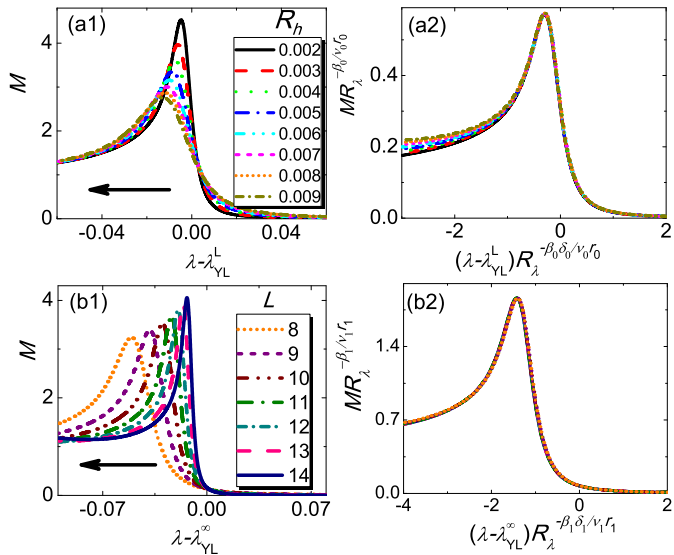,width=0.5\columnwidth}}
  \caption{\label{dbscalingl} (Color online) Under changing $\lambda$ with fixed $h=2.292475$, the curves of $M$ versus $(\lambda-\lambda_{\rm YL}^{L})$ ($\lambda_{\rm YL}^{L}=4.977885$) for fixed $L=10$ in (a1) match with each other in (a2) when $M$ and $(\lambda-\lambda_{\rm YL}^{L})$ are rescaled by the $(0+1)$D exponents; for comparison, the curves of $M$ versus $(\lambda-\lambda_{\rm YL}^\infty)$ for fixed $LR_\lambda^{1/r_1}=2.001689$ in (b1) match with each other in (b2) when $M$ and $(\lambda-\lambda_{\rm YL}^\infty)$ are rescaled by the $(1+1)$D exponents. The arrows point the directions of changing $\lambda$. $\lambda_0$ is chosen as $\lambda_0\equiv \lambda_{\rm YL}^{\infty}+1$ for both (a) and (b).}
\end{figure}

Then we study the relation between Eqs. (3) and (4) for changing $\lambda$. Similar to the procedure of changing $h$, we extract the order parameters at $\lambda_{\rm YL}^L$ for $R_\lambda$. Similar to Fig. 4(a) in the main text, Fig.~\ref{sclambda}(a) shows that for different lattice sizes, the curves of $M$ versus $R_\lambda$ are almost parallel lines on a double-logarithmic scale. By linearly fitting, we find that the slopes are between $-0.327$ and $-0.318$, close to the corresponding theoretical value for changing $h$. Similar to Fig. 4(b), we plot in Fig.~\ref{sclambda}(b) the rescaled order parameters as the function of the rescaled lattice sizes with the $(1+1)$D exponents for changing $h$ as input. We find that the rescaled curves collapse onto each other, confirming Eq.~(5) with $h$ and $R_h$ being replaced by $\lambda$ and $R_\lambda$, respectively. Moreover, linearly fitting the rescaled curve, one finds that the slope is about $-0.713$, agreeing with the theoretical value of $r_1(\beta_0\delta_0/\nu_0 r_0-\beta_1\delta_1/\nu_1 r_1)\simeq-0.7333$, confirming Eq.~(6) with $h$ and $R_h$ being replaced by $\lambda$ and $R_\lambda$, respectively.
\begin{figure}
  \centerline{\epsfig{file=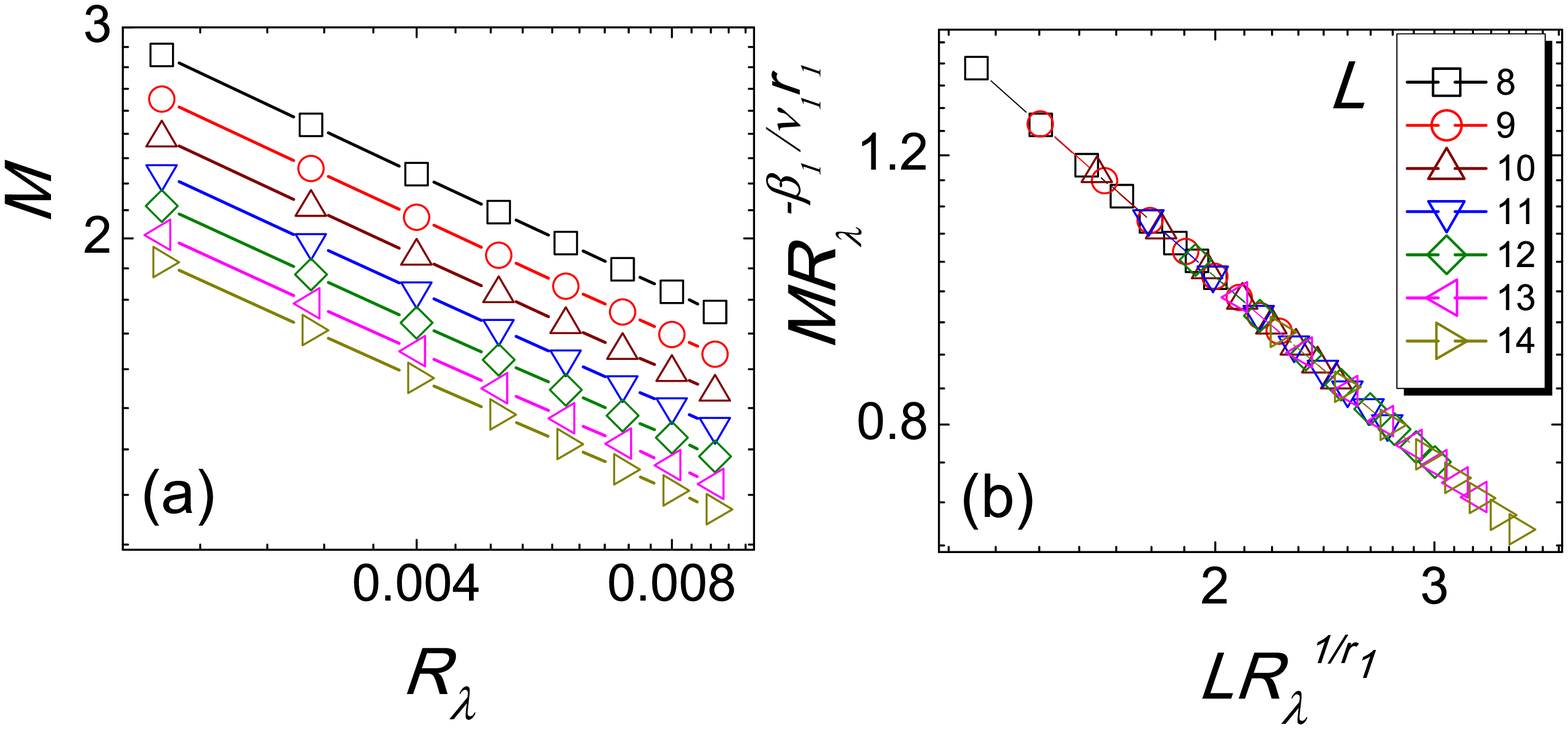,width=0.6\columnwidth}}
  \caption{\label{sclambda} (Color online) Under changing $\lambda$ with fixed $h=2.292475$, (a) $M$ at $\lambda_{\rm YL}^L$ versus $R_\lambda$ for different lattice size; and (b) the collapse of curves of the rescaled $M$ at $\lambda_{\rm YL}^L$ versus the rescaled $L$. Double-logrithmaic scales are used in both (a) and (b).}
\end{figure}

\subsection{\uppercase\expandafter{\romannumeral4}. The KZS for changing real longitudinal field $j$}
Here, we show that the KZS of the YLES is still applicable for the case of changing a real-longitudinal-field. The Hamiltonian reads
\begin{equation}
\mathcal{H}=-\sum_{n=1}^{L}\sigma_n^z\sigma_{n+1}^z-\lambda\sum_{n=1}^{L}\sigma_n ^x-{\rm i}h\sum_{n=1}^{L}\sigma_n^z-j\sum_{n=1}^{L}\sigma_n^z,
\label{HIsingsup}
\end{equation}
in which $j$ is the real-longitudinal field. The YLES occurs at $(\lambda_{\rm YL}^L, h_{\rm YL}^L)$ and $j=0$~\cite{Fishersup}. Here, the order parameter is calculated by $M={\rm Re}[\langle\Psi^*|\sigma^z|\Psi\rangle/\langle\Psi^*|\Psi\rangle]$ without the absolute-value operator, since the introduction of $j$ breaks the parity-time symmetry.

We consider the driven dynamics for changing $j$ as $j=j_0+R_jt$. Figure~\ref{dbscalingrhz}(a) show the curve of $M$ versus $j$ at $(\lambda_{\rm YL}^L, h_{\rm YL}^L)$ with a fixed $L$. After rescaling $M$ and $j$ by $R_j$ with the $(0+1)$D critical exponents, we find that the rescaled curves match with each other in the vicinity of $j=0$. Figure~\ref{dbscalingrhz}(b) shows the curve of $M$ versus $j$ at $(\lambda_{\rm YL}^\infty, h_{\rm YL}^\infty)$ with a fixed $LR_h^{1/r_1}$. After rescaling $M$ and $j$ with the $(1+1)$D exponents, we find that the rescaled curves collapse onto each other. These results show that similar to the case of changing the imaginary longitudinal field, the driven scaling in the overlap region can be described by both the $(0+1)$D and $(1+1)$D Kibble-Zurek scaling theories.
\begin{figure}[htb]
  \centerline{\epsfig{file=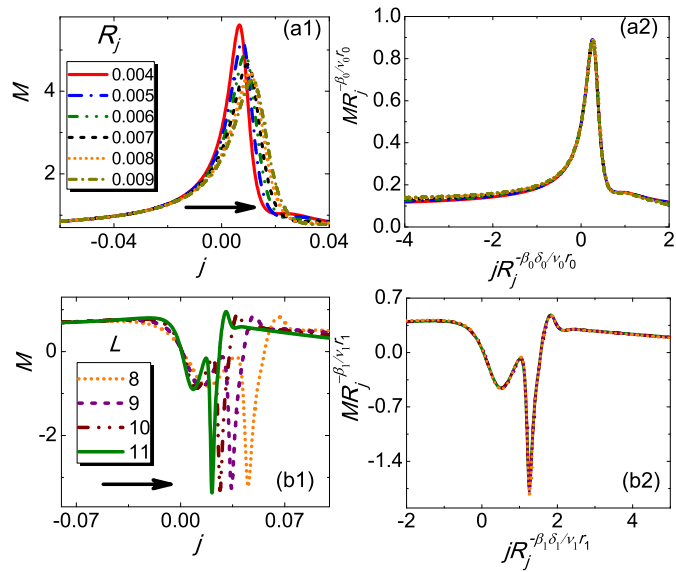,width=0.5\columnwidth}}
  \caption{\label{dbscalingrhz} (Color online) Under increasing $j$ at ($\lambda_{\rm YL}^{L}=5$, $h_{\rm YL}^{L}=2.3091759$) for fixed $L=10$, the curves of $M$ versus $j$ in (a1) match with each other in (a2) when $M$ and $j$ are rescaled by the $(0+1)$D exponents; for comparison, the curves of $M$ versus $j$ for fixed $LR_j^{1/r_1}=2.001689$ at ($\lambda_{\rm YL}^{L}=5$, $h_{\rm YL}^{L}=2.292475$) in (b1) match with each other in (b2) when $M$ and $j$ are rescaled by the $(1+1)$D exponents. $j_0$ is chosen to be $j_0=-1$.}
\end{figure}

\subsection{\uppercase\expandafter{\romannumeral5}. The KZS in the parity-time symmetry-breaking phase transition}
It has been shown that the YLES in model (1) in main text can be regarded as the critical phenomena of a parity-time symmetry-breaking phase transitions. Recently, a simple model, which exhibits the parity-time symmetry-breaking phase transition, has been realized in an experiment with ultracold atoms~\cite{LeLuo}. Therein an open quantum system is manipulated to simulate a parity-time symmetric system. The effective Hamiltonian for this experiment is
\begin{equation}
\mathcal{H}_{\rm eff}=-\lambda \sigma^x-{\rm i}h\sigma^z.
\label{HIsingsup1}
\end{equation}
Comparing Eq.~(\ref{HIsingsup1}) with model~(1) in main text, we find that Eq. (\ref{HIsingsup1}) describes the single particle limit of the quantum Ising model in an imaginary field and its transition belongs to the universality class of the $(0+1)$D YLES \cite{Griff}. So, the KZS in Eq.~(\ref{HIsingsup1}) should satisfy Eq.~(3) in the main text. The critical point for the parity-time symmetry-breaking is $h_{\rm YL}^L\equiv\lambda$ for $L=1$.

By changing $h$ linearly and calculating the order parameter $M$, we verify the Eq.~(3) in the main text for Eq.~(\ref{HIsingsup1}) in Figure \ref{onesite}. From Fig. \ref{onesite}(b), one finds that the curves of $M$ versus $h_{\rm YL}^L$ for different driving rates collapse onto each other perfectly after rescaling according to Eq.~(3) in the main text. We note that here the number of site is only one. So the mechanism for the emergent of the topological defects breaks down, as we discussed in the main text. In spite of this, the KZS is still applicable. Moreover, it has been shown that the dynamics for some open quantum systems can be described by the non-Hermitian Hamiltonian similar to Eq.~(1) in the main text~\cite{Schaller}. So, one can expect that our scaling theory can be observed in these systems.
\begin{figure}[htb]
  \centerline{\epsfig{file=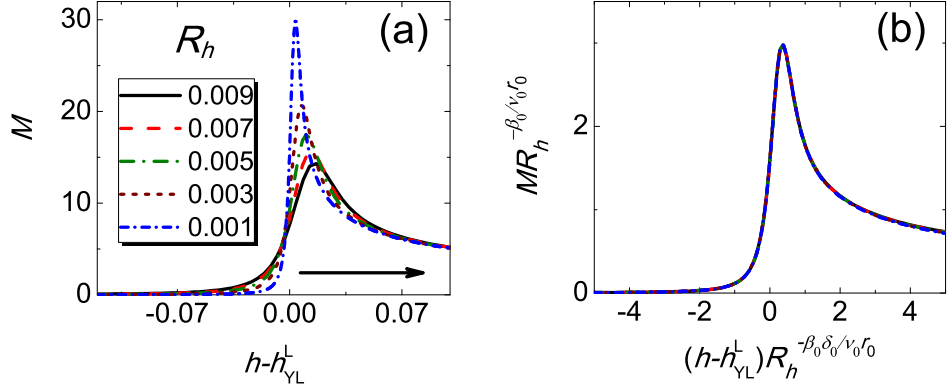,width=0.7\columnwidth}}
  \caption{\label{onesite} (Color online) Under increasing $h$ at $\lambda=5$ for fixed $L=1$, the curves of $M$ versus $h-h_{\rm YL}^{L}$ in (a) match with each other in (b) when $M$ and $h$ are rescaled by the $(0+1)$D exponents. $h_0$ is chosen to be $h_0=4$.}
\end{figure}

\end{widetext}

\end{document}